\documentclass[%
 aps,
 superscriptaddress,
 sd,%
 reprint,%
showpacs
]{revtex4-1}

\usepackage{amsmath,amssymb}
\usepackage{color}
\usepackage{graphicx}
\usepackage{dcolumn}
\usepackage{bm}

\begin{document}


\title[Topological insulator particles as optically induced oscillators: towards dynamical force measurements and optical rheology]{Topological insulator particles as optically induced oscillators: towards dynamical force measurements and optical rheology}

\author{W. H. Campos}
\author{J. M. Fonseca}
\affiliation{Departamento de
F\'isica, Universidade Federal de Vi\c{c}osa, 36570-900, Vi\c{c}osa, Minas Gerais, Brazil.}
\author{V. E. de Carvalho}
\affiliation{Departamento de F\' isica, ICEx, Universidade Federal de Minas Gerais, 30123-970, Belo Horizonte, Minas Gerais, Brazil.}
\author{J. B. S. Mendes}
\email{joaquimbsmendes@gmail.com}
\author{M. S. Rocha}
\author{W. A. Moura-Melo}
\affiliation{Departamento de
F\'isica, Universidade Federal de Vi\c{c}osa, 36570-900, Vi\c{c}osa, Minas Gerais, Brazil.}%

\keywords{topological insulators, optical tweezers, optical rheology, optically induced oscillators}

\date{\today}

\begin{abstract}

We report the first experimental study upon the optical trapping and manipulation of topological insulator (TI) particles. By virtue of the unique TI properties, which have a conducting surface and an insulating bulk, the particles present a peculiar behaviour in the presence of a single laser beam optical tweezers: they oscillate in a plane perpendicular to the direction of the laser propagation. In other words, TI particles behave as optically induced oscillators, allowing dynamical measurements with unprecedented simplicity and purely optical control. Actually, optical rheology of soft matter interfaces and biological membranes, as well as dynamical force measurements in macromolecules and biopolymers, may be quoted as feasible possibilities for the near future.\\

\end{abstract}


\maketitle

Since the seminal papers by Ashkin and collaborators \cite{Ashkin1,
Ashkin2}, the optical trapping and manipulation of micrometer-sized
particles have found applications in many areas, from interface and
colloid science to single molecule biophysics \cite{Svoboda, Wang,
Grier2}. In addition, improvements have been achieved regarding the use of optical tweezers as a tool for materials science \cite{Hansen, Reece, Bosanac, Juan}, and more recently, in the study of solution-phase chemistry \cite{Black}. Nowadays, a common optical tweezers setup consists of
a laser beam focused by a microscope objective of large numerical
aperture. This apparatus can trap dielectric objects near the lens
focus, being a powerful tool to capture and manipulate particles
with sizes at micrometer scale. The optical trapping of dielectric
particles are based on the competition between radiation pressure,
which occurs whenever light is reflected or absorbed by the
particles, and gradient force, coming about from light refraction
\cite{RochaAJP2}. In general, stable trapping occurs whenever
gradient force dominates, in such a way that the dielectric
particles are kept around the laser beam focus \cite{RochaAJP2}.
Furthermore, optical trapping demands dielectric particles with
refractive index higher than that of its surround medium (deionized
water, $n\approx 1.33$, is used in most experiments). On the other
hand, except for very special conditions, metallic particles cannot
be stably trapped by optical tweezers
\cite{Svoboda2, Min, ZhanQ, HuangL}.

In turn, topological insulators (TI's) are materials with unique
properties, whose robust stability is topologically protected by
time reversal symmetry. They are known to have insulating
(dielectric) bulk, but conducting surface states that support charges flowing without dissipation. These states are characterized by a
single gapless Dirac cone and exhibit the remarkable spin-momentum
locking: a charge carrier moves in such a way that its momentum is
always perpendicular to its spin \cite{reviewTI1, reviewTI2,
reviewTI3, reviewTI4}. By the way, curvature has been shown
to deeply affect the physical properties of TI's. For instance, it
is predicted that spherical \cite{spherical1, spherical2} or
toroidal TI \cite{toroidal} geometry renders a distinct
photo-emission spectrum. In addition, the electromagnetic response of
TI's is such that an external electric (and/or magnetic)
field induces both, magnetization and polarization as well, the so-called
topological magnetoelectric effect. As a direct consequence,
whenever light comes onto a TI, reflected and refracted rays
experience topological Faraday and Kerr rotations, respectively
\cite{reviewTI1, reviewTI2, reviewTI3, reviewTI4}. Furthermore, it
has been recently predicted that the topological Kerr effect gives
rise to a residual force perpendicular to the incident plane
whenever light is shed onto a magnetically capped spherical topological insulator bead
\cite{yuri}. Such a topological-like force goes around some dozens
of femtoNewtons, about the Casimir force scale
\cite{Casimir-force-EPL}. Thus, whenever subjected to a highly
focused light beam, like those used in optical tweezers, it is
expected that a TI particle should experience competing effects
coming from the interaction of light with its conducting surface and
insulating bulk. We should wonder about the resulting effect of such an
interaction to the TI-particle dynamics.

In this work, we conduct an experimental study
regarding the optical trapping and manipulation of TI particles. Even though our present investigation was initially motivated by the prediction reported in Ref.\cite{yuri}, we did not expect to carry out a direct observation of such a topological-like force, mainly for we have used TI-particles without any magnetic cap, whose field would open an energy gap at the surface states, making topological Kerr effect possible. Therefore, our results reported in this Letter do not confirm such a theoretical prediction (nor do rule it out). Actually, by
virtue of their special characteristics, these particles present a
quite unusual dynamics under a highly focused light beam: they
oscillate perpendicularly to the direction of laser propagation. In
other words, they behave as optically induced oscillators, making them unique candidates to open an avenue for novel applications of
optical manipulation techniques, allowing dynamical measurements
with unprecedented simplicity. Actually, rheology of soft matter
interfaces and biological membranes, dynamical force
measurements in macromolecules and biopolymers may be quoted as some of feasible achievements by using TI-beads as optical oscillators.

The particles of Bi$_2$Te$_3$ and Bi$_2$Se$_3$ are topological insulators at room temperature and have been obtained by laser ablation technique (see details in the Supplementary
Material \cite{supmat}). Measurements were made for both composites and they present the same general behaviour. In order to be more succinct, we present here only the results for Bi$_2$Te$_3$, which have a more spherical shape than the Bi$_2$Se$_3$ ones. For the optical experiment, we have
selected only those particles with a nearly spherical shape, and average diameter between 3 $\mu$m and 7 $\mu$m. Later, they have
been suspended in deionized water and placed in the sample chamber,
which consists of an {\em o-ring} glued in a microscope coverslip. In fact, we have observed that the beads remain suspended around $4\, \mu$m above the coverslip surface, before and during the laser incidence. The optical tweezers consist of a 1064 nm ytterbium-doped fiber laser
(IPG Photonics) operating in the TEM$_{00}$ mode, mounted on a Nikon
Ti-S inverted microscope with a 100$\times$ NA 1.4 objective. In all
experiments we have used a laser power of 25 mW measured at the
objective entrance.

\begin{figure}
\centering
\includegraphics[width=7.5cm]{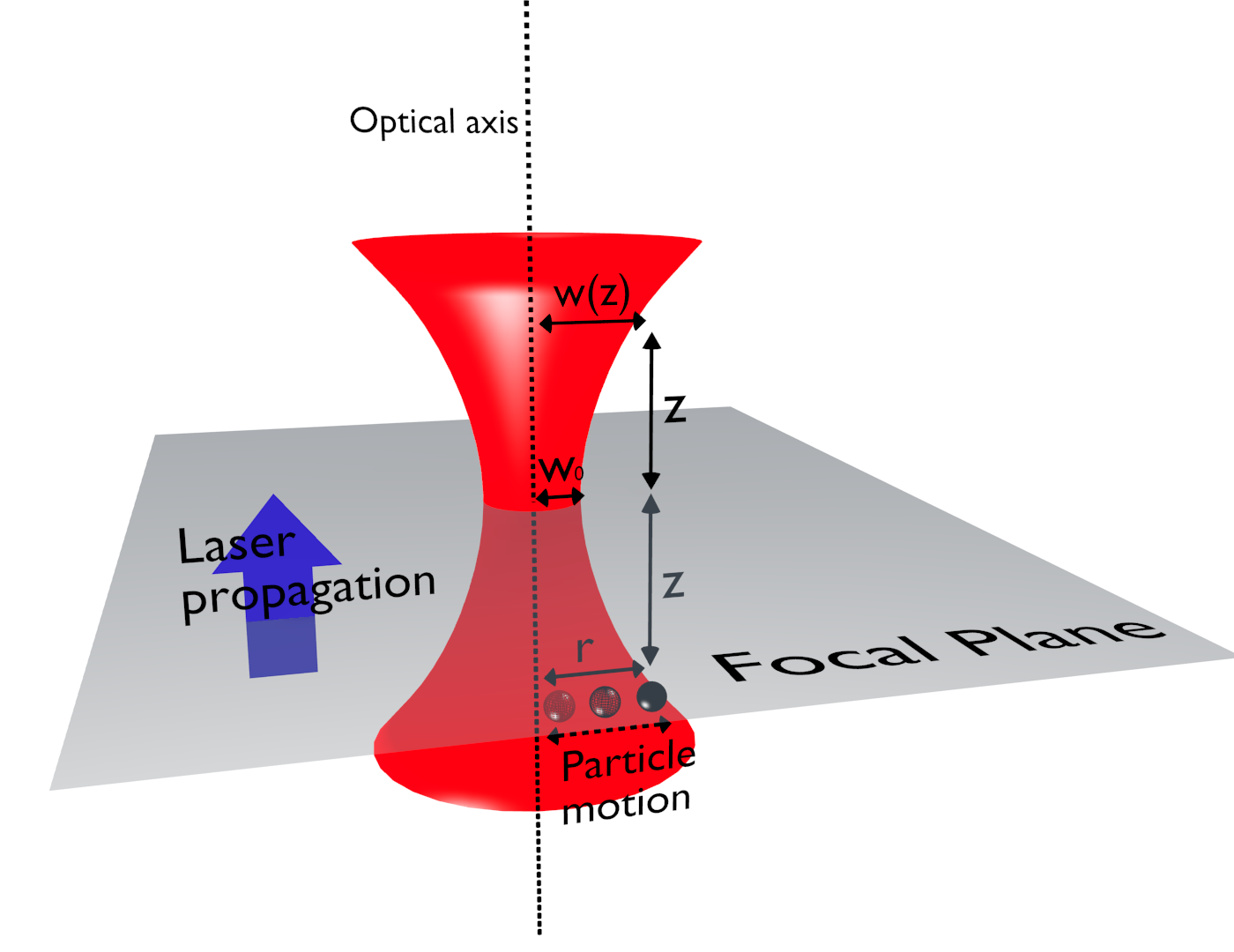}
\caption{(Color online) Experimental setup yielding the oscillatory dynamics and the relevant parameters. A TI particle is located around $z \sim$ 10 $\mu$m below the focal plane and it oscillates perpendicularly to
the optical axis direction. The particle lies within the optical beam waist, i.e. $r(t) < w(z)$.} \label{twzgeometry}
\end{figure}

Fig. \ref{twzgeometry} shows our experimental setup along with the
relevant parameters. TI particles located $z \sim$ 10 $\mu$m below
the focal plane have been observed to oscillate perpendicularly to the optical axis
direction, say, parallel to the focal plane. On the other
hand, if the particle is close enough to the focal plane ($z
\lesssim$ 3.5 $\mu{\rm m}$), then the resultant repulsive force becomes high enough to drift it away, like occurs to typical metallic particles.
Here we propose a simple model that takes into account the
competition between gradient forces and the resulting repulsive force that acts on the TI particles. Such repulsive force is mainly due to the radiometric effect that plays a role when the particles considerably absorb light from the laser beam (radiometric force). Our model allows one to predict the resultant force on the particle yielding its oscillatory dynamics (details upon the model below and how it is concerned with the characteristics of topological insulator, namely, its surface density of states, may be found in the Supplementary Material \cite{supmat}). At a
certain vertical distance $z$ from the focal plane ($z$ = 0), the
laser intensity at an arbitrary position ($r$, $z$), normalized by
the corresponding intensity at the optical axis ($r$ = 0, $z$), can
be written appropriately, for a Gaussian beam profile, by:

\begin{equation}
I_N = \exp{\left(\frac{-2r^2}{w(z)^2}\right)} \label{intprofile},
\end{equation}
where $w(z)$ is the beam waist at position $z$; it is related to the beam waist at the focal
plane, $w$($z$ = 0) $\equiv w_0$, as below:

\begin{equation}
w(z) = w_0\sqrt{1 + \left(\frac{z}{z_R} \right)^2} \label{wz}, \qquad z_R = \frac{\pi w_0^2}{\lambda}.
\end{equation}
Whenever we focus our attention at a height where the particle has no dynamics along the laser beam direction, then we are only left with non-vanishing force in the plane perpendicular to the optical axis. In addition, let $r$ label the direction along with the particle dynamics is predominant, as observed in our experiments. Therefore, we are left with the problem of determining the net force dictating the particle dynamics along $r$, as it follows. The effective resulting radiometric force acting on the particle is proportional to the laser intensity reaching it \cite{Ke}. Therefore we write
\begin{equation}
F_{r} = \mathcal{F}_{r}\exp{\left(\frac{-2r^2}{w(z)^2}\right)}
\label{FRP},
\end{equation}
where $\mathcal{F}_{r}$ accounts for the maximum magnitude of this force, at $r$ = 0. Observe that equation (\ref{FRP}) also implicitly includes the resulting effect of radiation pressure on the particle, since this force is also proportional to the laser intensity.

In turn, the gradient force, is given by:
\begin{equation}
F_{g} = - \frac{4rA}{w(z)^2}\exp{\left(\frac{-2r^2}{w(z)^2}\right)}
\label{FG},
\end{equation}
since $F_g$ is proportional to the negative derivative of the beam
intensity. The constant $A$ is proportional to the maximum magnitude of the
gradient force, $\mathcal{F}_{g}$, which occurs at $r =
(1/2)w$($z$), by:
\begin{equation}
A = \frac12\mathcal{F}_{g}w(z)\exp(1/2)\label{A}.
\end{equation}

Finally, the resultant force $F$ = $F_{r}$ + $F_{g}$ can be
written as:

\begin{equation}
F = \left(\mathcal{F}_{r} -
\frac{2r\mathcal{F}_{g}\exp(1/2)}{w(z)}\right)\exp{\left(\frac{-2r^2}{w(z)^2}\right)}
\label{Fres}.
\end{equation}

\begin{figure}
\centering
\includegraphics[width=7.5cm]{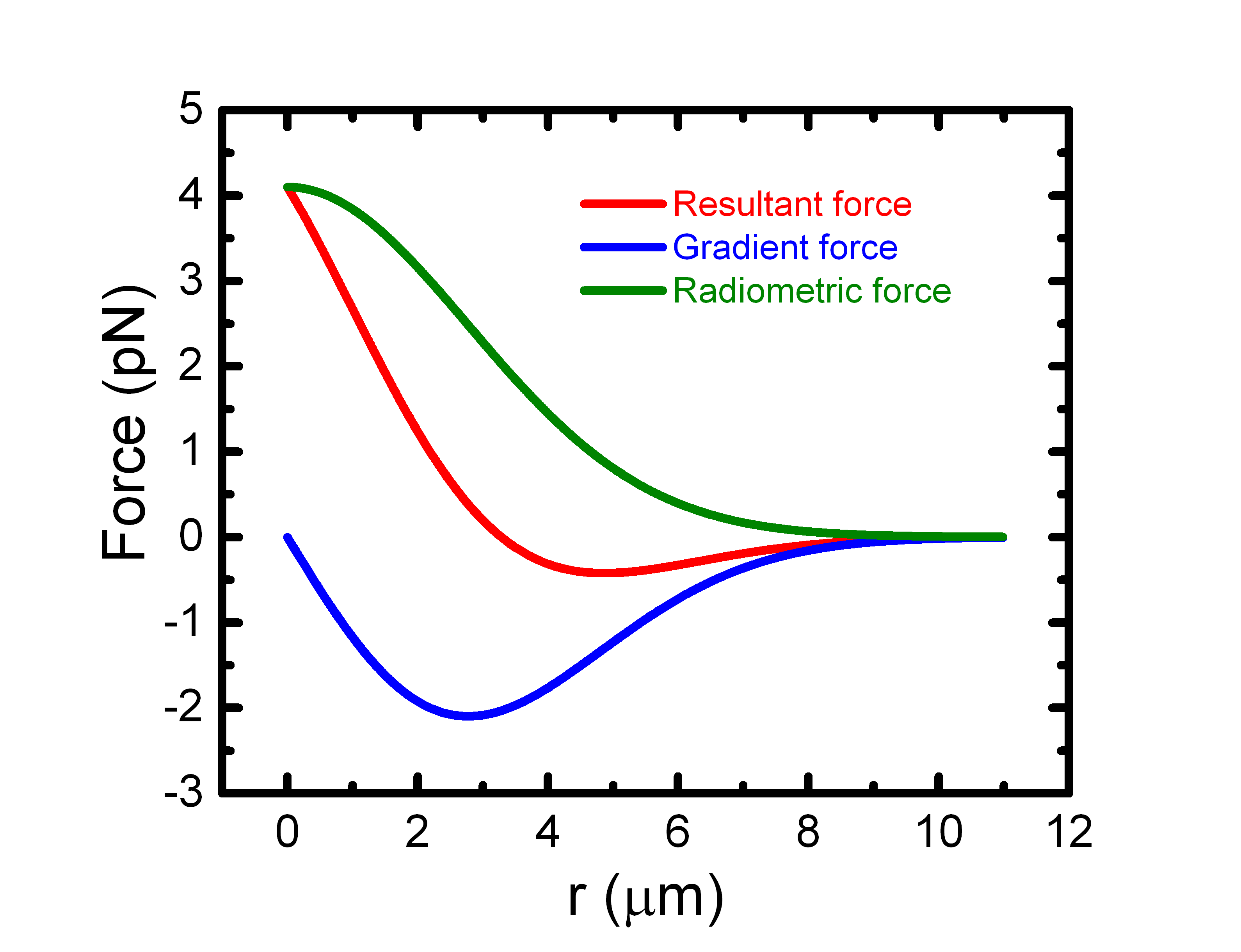}
\caption{(Color online) Typical theoretical behaviour of the forces as functions of
TI-particle position, $r$. Here, we have used the parameters
$w$($z$) = 5 $\mu$m, $\mathcal{F}_{r}$ = 4 pN, and $A$ = 10
pN.$\mu$m, corresponding to $\mathcal{F}_{g}$ $\sim$ 2.43 pN.}
\label{opticalforces}
\end{figure}

In Fig. \ref{opticalforces} it is shown how such forces behave as
$r$ varies. For that, we have taken: $w$($z$) = 5 $\mu$m,
$\mathcal{F}_{r}$ = 4 pN and $A$ = 10 pN.$\mu$m, corresponding to
$\mathcal{F}_{g}$ $\sim$ 2.43 pN. For these parameters the resultant force is repulsive for $r <$ 2.5 $\mu$m and attractive for
$r >$ 2.5 $\mu$m. Indeed, the value of $r$ where the crossover
between repulsive and attractive regimes occurs is given by $r_c$ =
$\mathcal{F}_{r}$$w$($z$)/2$\mathcal{F}_{g}\exp(1/2)$. In addition,
as long as $\mathcal{F}_{r}$/2$\mathcal{F}_{g}\exp(1/2)$ $>$ 1,
then $r_c >$ $w$($z$), yielding the repulsive regime, say,
radiometric force dominates over gradient force and no oscillatory
motion takes place. At this point, it should be stressed that metallic
beads experiences only the repulsive regime, once the radiometric forces (and also radiation pressure) are much higher than gradient force. Conversely, whenever ordinary dielectric particles are in order, gradient force largely dominates, yielding the usually observed trapping around the optical focus. Our findings regarding TI microsized beads open a new possibility, once they behave like light induced oscillators, whose quasiperiodic motion may be used to probe important properties in a number of experiments concerning optical tweezers, namely dynamical force measurements and microrheology studies.

The procedure used to obtain the forces in the experiments
is detailed in the Supplementary Material \cite{supmat}. Essentially, the
bead position is recorded using videomicroscopy, from which other dynamical variables, like velocity and acceleration are readily obtained, as well as, the net force acting on the
bead. Such a force comes from three independent contributions: the
optical, the radiometric, and the viscous (Stokes) force. Actually, in order to obtain the resultant force
for comparison with the theoretical model described above, equation (\ref{Fres}), one needs to subtract the contribution coming from the viscous force, which was calculated using the instantaneous velocity of the particle. [Three movies showing the motion of
Bi$_2$Te$_3$ TI-particles can also be found as a Supplementary
Material \cite{supmat}].

\begin{figure}
\centering
\includegraphics[width=7.5cm]{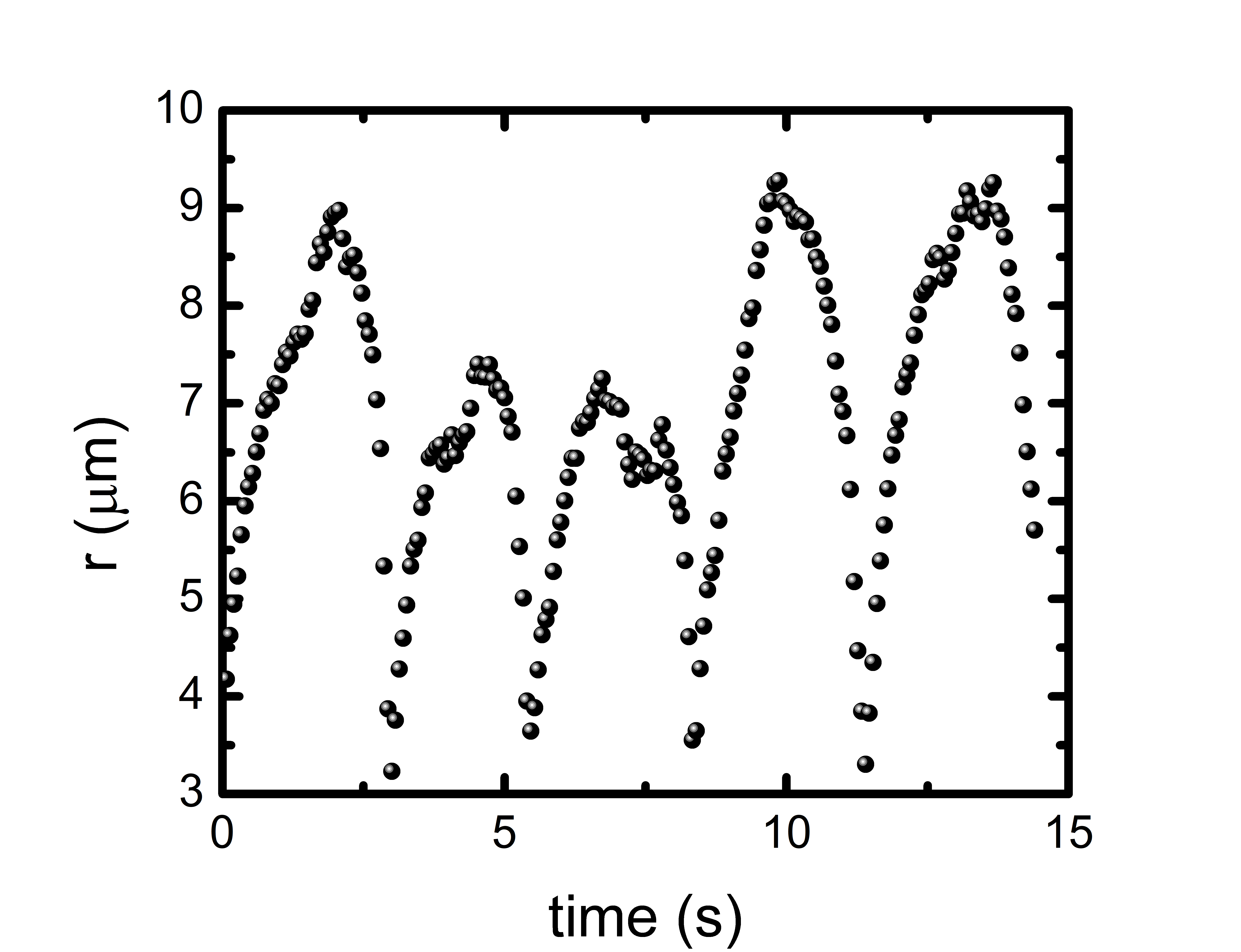}
\caption{(Color online) Typical dynamics of the TI-particles, $r(t)\times t$. The particle position $r(t)$ is measured relative to the optical axis, at $r=0$. These experimental data have been obtained for
a single Bi$_2$Te$_3$ spherical-like bead with average
diameter $\sim$ 4.2 $\mu$m. The bead have been placed at $z \sim$ 10
$\mu$m. The motion presents a well-defined frequency for the
oscillations, with period $T\approx 3$ ${\rm s}$. The uncertainty in the measurement of position $r(t)$ is $0.017$ $\mu$m.} \label{oscillations}
\end{figure}

\begin{figure}
\centering
\includegraphics[width=7.5cm]{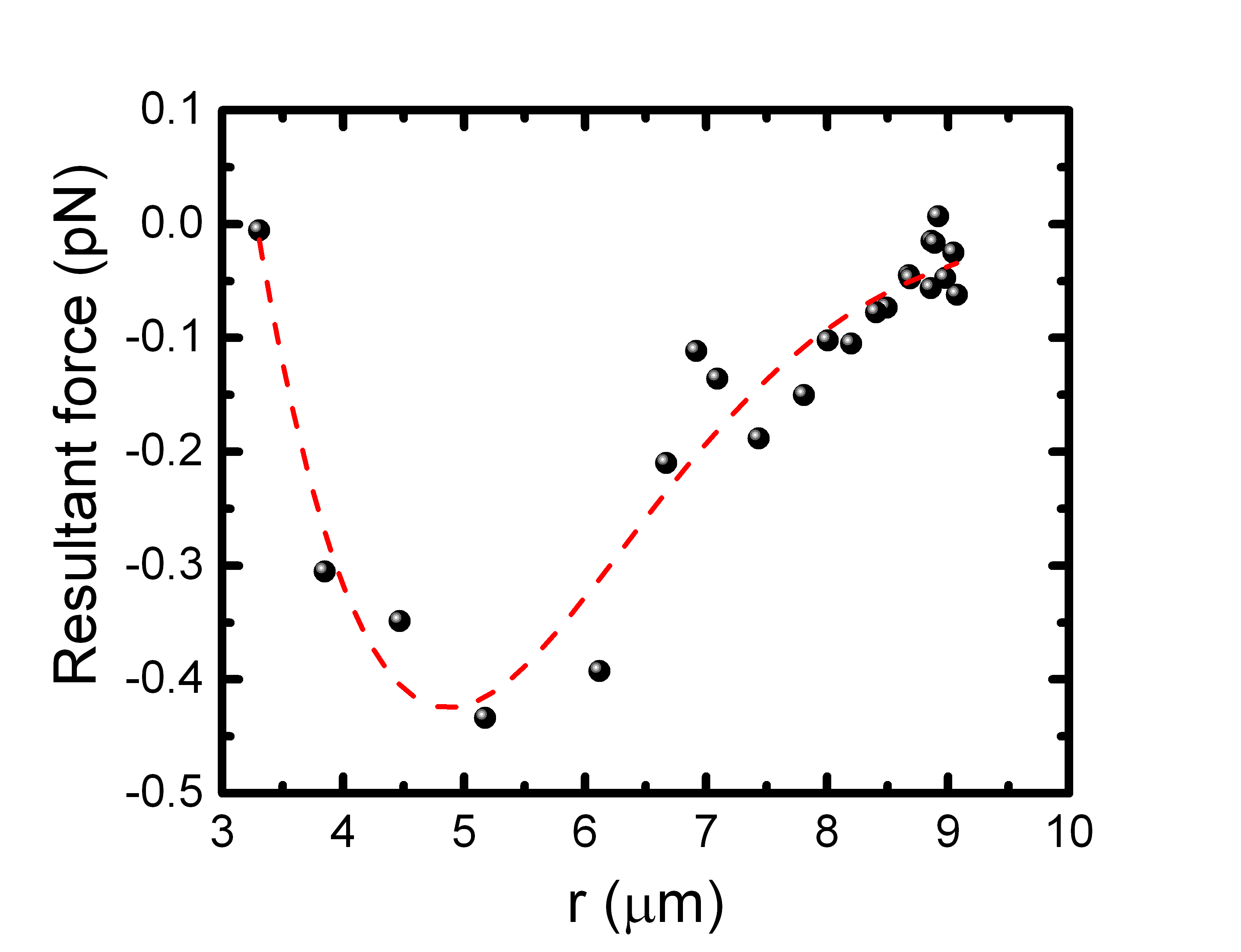}
\caption{(Color online) Resultant force as function of particle
position $\mathbf{r}$. Here, the particle is moving towards the optical axis
(attractive regime). Experimental data (\textit{black circles}) are
well-fitted by the theoretical model (\textit{dashed line}), equation
(\ref{Fres}). The estimated mean uncertainty in the calculation of the force is $0.03$ pN. } \label{fitapproach}
\end{figure}

In Fig. \ref{oscillations} it is shown the typical dynamics of the
TI-particles, where its position relative to the optical axis, $r$, is plotted as function of the time, $t$. Such results have been obtained for a Bi$_2$Te$_3$ spherical-like bead with diameter $\sim$ 4.2 $\mu$m
centered at $z \sim$ 10 $\mu$m. The oscillations are well-defined in
time with period $T\approx 3$ ${\rm s}$. Their amplitudes vary between
$\sim 7$ $\mu{\rm m} \,-\, 9$ $\mu{\rm m}$, with closest approximation to
the optical axis observed to be $\sim 3.2$ $\mu{\rm m}$. In
addition, the variation in the amplitudes comes from the
fact that $r(t)$ only computes the motion parallel to the focal
plane, whereas the particle position sometimes presents small
fluctuations along $z$ direction. Besides, once the  TI particles are not perfect spheres, some deviations in the amplitude could be also associated to their shape. 
Fig. \ref{fitapproach} shows the resultant force (gradient force + radiometric force) as a function of the particle position $r$ for the attractive regime
(particle approaching the optical axis). These example data have been obtained from the first oscillatory cycle shown in Fig.
\ref{oscillations} (details in the Supplementary Material \cite{supmat}).
In addition, it is noteworthy that our model fits the experimental
data accurately, with the following values for the physical
parameters: $w$($z$) = (5.55 $\pm$ 0.15) $\mu$m, $\mathcal{F}_{r}$
= (4.1 $\pm$ 0.6) pN and $\mathcal{F}_{g}$ = (2.1 $\pm$ 0.2) pN. For the other oscillatory cycles, the variations in the oscillation amplitude impart on the values of the forces $\mathcal{F}_{r}$ and $\mathcal{F}_{g}$, making them to vary from one cycle to the other, namely whenever very distinct amplitude cycles are considered. On the
other hand, it is worthy to mention that the parameter $w$($z$) is
very robust against amplitude variation, what is expected since
$w$($z$) is a characteristic of the laser beam. Its average value,
found considering different particle motions, reads $w$($z$) = (5.7
$\pm$ 0.3) $\mu$m. Taking $z \sim$ 10 $\mu$m to equation (\ref{wz}), yields
$w_{0_{exp}} = (0.45 \pm 0.02)$ $\mu$m, which is in good agreement
to the predicted value 2$\lambda$/$\pi$$NA$ $\sim$ 0.36 $\mu$m
\cite{Kiang}. The
repulsive regime, say, particle moving away from the optical axis, is
discussed in the Supplementary Material \cite{supmat}.

\begin{figure}
\centering
\includegraphics[width=7.5cm]{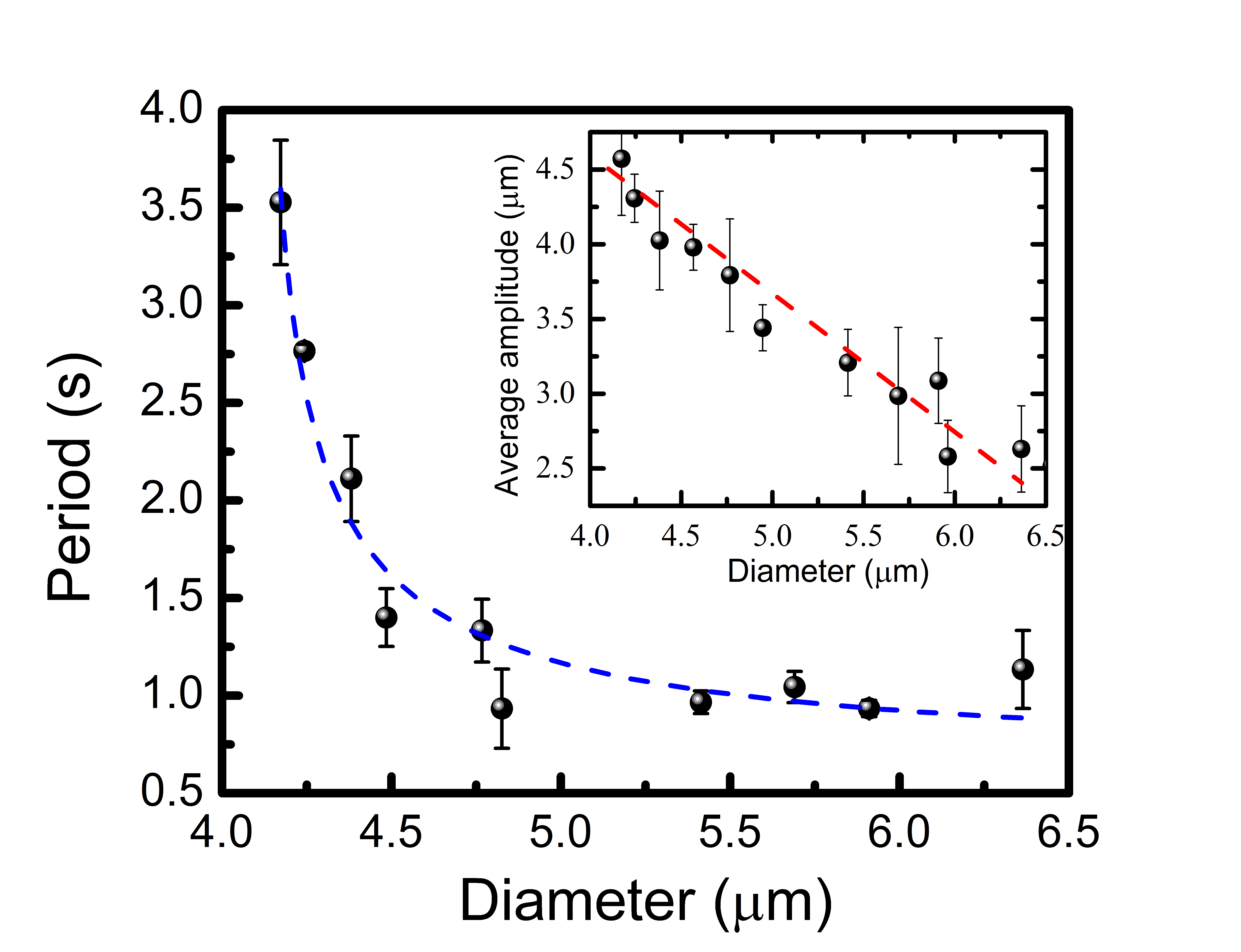}
\caption{(Color online) Period of the oscillatory motion as a function of the
particle diameter. Experimental results (black circles) along with their respective errors (black bars). Note that as particle size increases the period diminishes in a way relatively well-described within the harmonic oscillatory regime (traced fitting curve). The inset shows the average amplitude for the particles with different sizes. Dashed red line is just a guide for the eyes and it illustrates its linear decreasing with diameter. All these measurements have been conducted with the particles around $z= 10\,\mu$m, where oscillations show up more evident.} \label{frequency}
\end{figure}

Fig. \ref{frequency} shows how the period of the oscillations depends upon beads diameter. Smaller beads tend to oscillate slower than larger ones. For instance, a bead with diameter $\sim 4.2$ $\mu{\rm m}$ has a period $\sim 3.5$ ${\rm s}$ (frequency $\sim 0.3$ ${\rm Hz}$), while another with $\sim 6$ $\mu{\rm m}$ has its frequency increased to $\sim 1$ ${\rm Hz}$. Although these oscillations are not harmonic (recall the expression of the force, equation (\ref{Fres}), along with the viscous dissipation), such a behaviour is well-captured within the simple harmonic regime, as follows. For that, recall that the gradient force comes from volumetric bulk refractions, then as particle radius, $a$, varies it is expected that $F_g \sim A a^3$. Once radiometric force increases with the particle area then $F_r \sim B a^2$, while Stokes force goes like $F_S\sim C a$, where $A,B,$ and $C$ are constants. Restricting ourselves to simple harmonic oscillatory description dictated by these forces, the period, $T\sim\sqrt{m/k}$ (spherical particle mass $m=4\pi\rho a^3/3$) goes like $T \sim \sqrt{\frac{a^3}{A a^3 + Ba^2 +C a}}$. For the fitting depicted in Fig. \ref{frequency} we have $T' \sim \sqrt{\frac{a^3}{A' a^3 + B'a^2 +C' a}}$, with A$'$ $\sim$ 1.81 s$^{-2}$, B$' \sim$ 3.89 $\mu$ms$^{-2}$, and C$'\sim$ -46.50 $\mu$m$^2$ s$^{-2}$.\\

An important point to be stressed here concerns the actual possibility of tunning the particle period/frequency by just varying its size. Other improvements are certainly obtained by adjusting other physical parameters, like the beam power and so forth. In fact, we have observed that both amplitude and period of oscillations increase with laser power (see discussion in the Supplementary Material \cite{supmat}). One may wonder whether our observations regarding TI optically induced oscillators may occur to other materials-made microspheres. To our best findings, this may be the case provided that the beads have an intermediary surface conductivity (density of charged states) along with a good transparent bulk (relatively high skin depth). These two key characteristics are accomplished by topological insulator beads, as discussed in the Supplementary Material \cite{supmat}. Whether other materials may present similar characteristics remains to be investigated and it should enlarge even more the issue of optically induced oscillators. 
%

\vskip 0.5cm

In summary, we have observed that microsized topological insulator (Bi$_2$Te$_3$ and Bi$_2$Se$_3$) particles oscillate perpendicularly to the optical axis  whenever subject to a highly focused light beam . Physically, such oscillations are a result of a delicate balance between gradient forces and radiometric forces. This last component acts as a driving force, changing the usual overdamped motion obtained for ordinary dielectric particles into a driven (but also damped) oscillatory motion. Even though the oscillations are not harmonic, the period/frequency appears to  remain practically constant during a number of cycles. Such a frequency is also dependent on the particle size, increasing with the beads diameter in such a way well-described by a simple harmonic approach. Both of these features are important for practical purposes and they may be further improved by combining particle size/shape with modulated intensity/power laser beam. For instance, beads having more regular spherical shape are crucial for highly precise experiments. This precision is important to make such optically induced oscillators useful in dynamical force measurements in macromolecules and biopolymers. They may also play an important role in optical rheology of biological membranes and soft matter interfaces.\\

\section{METHODS}
A detailed discussion upon the methodology used to prepare the samples and obtain the data may be found in the Supplementary Material \cite{supmat}.

\section{ASSOCIATED CONTENT}
\textbf{Supporting information}\\
The Supporting Information is available at DOI: \textcolor{blue}{[URL will be inserted by the publisher]}.\\
Additional text with details about the experimental procedure used to obtain the data, model relating the forces to topological insulator properties, the preparation and characterization of Bi$_2$Te$_3$ and Bi$_2$Se$_3$ crystals. (PDF)\\
Video-1 and Video-2: Real time videos showing the standard behaviour of topological insulators microparticles at our optical tweezers setup. (AVI)\\
Video-3: Real time video, in which we purposely used a dielectric particle in order to show its trapping, while the surrounding TI particles oscillate towards the optical axis. (AVI)

\section{AUTHOR INFORMATION}
\textbf{Corresponding Author}\\
*E-mail: joaquimbsmendes@gmail.com\\
\textbf{ORCID}\\
Warlley H. Campos: 0000-0002-4191-3020\\
Jakson M. Fonseca: 0000-0001-9441-026X\\
Vagner E. de Carvalho: 0000-0002-2357-9356\\
Joaquim B. S. Mendes: 0000-0001-9381-0448\\
M\' arcio S. Rocha: 0000-0003-0323-3718\\
Winder A. Moura-Melo: 0000-0003-3027-6852\\
\textbf{Author Contributions}\\
J.B.S.M. and V.E.C. performed the preparation and characterization of the samples. W.H.C., J.B.S.M. and M.S.R. performed optical tweezers experiments. W.H.C., J.M.F., J.B.S.M., M.S.R. and W.A.M-M. designed the experiment, analyzed the data and interpreted the results. All authors contributed to the writing of the manuscript and the supplementary material.\\
\textbf{Notes}\\
The authors declare no competing financial interest.

\begin{acknowledgements}
The authors would like to thank Professor Cid B. de Ara\' ujo (Department of Physics - Universidade Federal de Pernambuco) for reviewing the manuscript. They also express their gratitude to  CAPES, CNPq, FAPEMIG and FINEP (Brazilian agencies).
\end{acknowledgements}

\bibliography{Rocha2015_bibtex}

\end{document}